\documentclass[12pt]{article}
\pdfoutput=1
\usepackage{amsmath}
\def\be{\begin{equation}}
\def\ee{\end{equation}}
\def\arr{\begin{array}{rll}}
\def\ea{\end{array}}
\def\bea{\begin{eqnarray}}
\def\eea{\end{eqnarray}}

\def\N2{$N{=}2$}

\def\>{\rangle}
\def\<{\langle}
\def\+{\dagger}
\def\={\ =\ }

\begin{document}
\renewcommand{\thefootnote}{\fnsymbol{footnote}}
\begin{titlepage}
\setcounter{page}{0}
\vskip 1cm
\begin{center}
{\LARGE\bf Super--Schwarzians via nonlinear realizations}\\

\vskip 1.5cm
$
\textrm{\Large Anton Galajinsky\ }
$
\vskip 0.7cm
{\it
Tomsk State University of Control Systems and Radioelectronics, 634050 Tomsk, Lenin Ave. 40, Russia} \\
\vskip 0.2cm
{e-mail: anton.galajinsky@gmail.com}
\vskip 0.5cm

\end{center}
\vskip 1cm
\begin{abstract} \noindent
The $\mathcal{N}=1$ and $\mathcal{N}=2$ super--Schwarzian derivatives were originally introduced by physicists when computing a finite superconformal transformation of the super stress--energy tensor underlying a superconformal field theory. Mathematicians like to think of them as the cocycles describing central extensions of Lie superalgebras.
In this work, a third possibility is discussed which consists in applying the method of nonlinear realizations to $osp(1|2)$ and $su(1,1|1)$ superconformal algebras. It is demonstrated that the super--Schwarzians arise quite naturally, if one decides to keep the number of independent Goldstone superfields to a minimum.
\end{abstract}


\end{titlepage}

\renewcommand{\thefootnote}{\arabic{footnote}}
\setcounter{footnote}0

\noindent
{\bf 1. Introduction}\\

\noindent
A recent study of supersymmetric extensions of the Sachdev--Ye--Kitaev model \cite{FGMS} generated renewed interest in the $\mathcal{N}=1$ and $\mathcal{N}=2$ super--Schwarzian derivatives.\footnote{The literature on the subject is rather extensive. For a good recent account and further references see Ref. \cite{BBN}.}
Such derivatives were originally introduced by physicists when computing a (finite) superconformal transformation of the super stress--energy tensor underlying a superconformal field theory \cite{F,Cohn,Sch}.
Mathematicians used to regard them as the cocycles describing central extensions of Lie superalgebras (see, e.g., \cite{MD} and references therein).

A remarkable property of the $\mathcal{N}=1$ and $\mathcal{N}=2$ super--Schwarzian derivatives is that they hold invariant under (finite) transformations forming $OSp(1|2)$ and $SU(1,1|1)$ superconformal groups, respectively.\footnote{In modern literature, the superconformal groups are sometimes designated by the number of spacetime dimensions in which they are realised and the number of {\it real} supersymmetry charges at hand. In this nomenclature $OSp(1|2)$ and $SU(1,1|1)$ are identified with the $d=1$, $\mathcal{N}=1$ and $d=1$, $\mathcal{N}=2$ superconformal groups, respectively.} It is then natural to wonder whether the logic can be turned around so as to derive the super--Schwarzians by analysing invariants of the supergroups alone.

Given a Lie (super)algebra, a conventional means of building invariants associated with the corresponding (super)group is to apply the method of nonlinear realizations \cite{CWZ}. Within this framework,
one starts with a coset space element $\tilde g$, on which a (super)group representative $g$ acts by the left multiplication $\tilde g'=g \cdot \tilde g$, and then constructs the Maurer--Cartan one--forms
${\tilde g}^{-1} d \tilde g$, which are automatically invariant under the transformation. These invariants can be used to impose constraints enabling one to express some of the (super)fields parametrizing the coset element $\tilde g$ in terms of the other \cite{IO}. If the algebra at hand is such that all but one (super)fields can be linked to a single unconstrained (super)field, then the last remaining Maurer--Cartan invariant describes a derivative of the latter, which holds invariant under the action of the (super)group one started with.

As an illustration, let us consider the coset space element $\tilde g=e^{i\rho(t)P} e^{i s(t) K} e^{i u(t) D}$, which builds upon the generators $P$, $D$, $K$ forming $sl(2,R)$ algebra\footnote{$P$, $D$, and $K$ are associated with translations, dilatations, and special conformal transformations, respectively, and obey the structure relations
$[P,D]=i P$, $[P,K]=2 i D$, $[D,K]=i K$.} and three real functions $\rho(t)$, $s(t)$, $u(t)$ of a temporal variable $t$, and compute the Maurer--Cartan invariants $\tilde g^{-1} d \tilde g=i\left(\omega_P P +\omega_K K+\omega_D D\right)$ (see Ref. \cite{G1} for more details)
\be
\omega_P=\dot\rho e^{-u} dt, \qquad \omega_D=\left(\dot u-2 s \dot\rho\right) dt, \qquad \omega_K=e^u \left(\dot s+s^2 \dot\rho\right)dt.
\nonumber
\ee
The first two forms can be used to impose constraints $\dot\rho e^{-u}=\frac 12$, $\dot u-2 s \dot\rho=0$,
which link $u$ and $s$ to a single unconstrained $\rho$. Substituting the result into the last remaining invariant $e^u \left(\dot s+s^2 \dot\rho\right)$, one gets
\be
\frac{\dddot{\rho}(t)}{\dot\rho(t)}-\frac 32 {\left(\frac{\ddot{\rho}(t)}{\dot\rho(t)}\right)}^2.
\nonumber
\ee
This is the celebrated Schwarzian derivative. Note that within this framework the $SL(2,R)$--invariance of the latter is obvious as it is built from the Maurer--Cartan invariants.

The goal of this paper is to provide a similar derivation of the $\mathcal{N}=1$ and $\mathcal{N}=2$ super--Schwarzian derivatives.

The work is organized as follows. In the next section, the $\mathcal{N}=1$ super--Schwarzian derivative is obtained by applying the method of nonlinear realizations to $osp(1|2)$ superconformal algebra. Five superfield invariants are constructed which enter the decomposition of ${\tilde g}^{-1} \mathcal{D} \tilde g$, where $\mathcal{D}$ is the covariant derivative, into a linear combination of the generators of $osp(1|2)$.
Note that the Grassmann parity of ${\tilde g}^{-1} \mathcal{D} \tilde g$ is opposite to that of the conventional Maurer--Cartan invariant $\tilde g^{-1} d \tilde g$ because $\mathcal{D}$ is an odd operator.
Imposing four constraints so as to express four superfields parametrizing a coset space element $\tilde g$ in terms of one unconstrained fermionic superfield and substituting the result into the last remaining invariant, one reproduces the $\mathcal{N}=1$ super--Schwarzian derivative. A similar group--theoretic derivation of the $\mathcal{N}=2$ super--Schwarzian based upon $su(1,1|1)$ superalgebra is given in Sec. 3. In contrast to the previous case, the $\mathcal{N}=2$ super--Schwarzian derivative comes about when one analyses the reality condition for the superfield associated with the generator of special conformal transformations. We summarise our results and discuss possible further developments in the concluding Sect. 4. Some useful identities relevant for computation of the superconformal invariants in Sect. 2 and Sect. 3 are gathered in Appendix.

\vspace{0.5cm}

\noindent
{\bf 2. $\mathcal{N}=1$ super--Schwarzian derivative via nonlinear realizations}\\

\noindent
The $\mathcal{N}=1$ super--Schwarzian derivative
\be\label{ssd}
S[\psi(t,\theta);t,\theta]=\frac{\mathcal{D}^4 \psi}{{\mathcal{D}\psi}}-2 \frac{\mathcal{D}^3 \psi}{{\mathcal{D}\psi}} \frac{\mathcal{D}^2 \psi}{{\mathcal{D}\psi}}
\ee
where $\psi(t,\theta)$ is a real fermionic superfield and $\mathcal{D}$ is the covariant derivative,
was first introduced in \cite{F} by computing a finite superconformal transformation of the super stress--energy tensor underlying an $\mathcal{N}=1$ superconformal field theory.\footnote{To be more precise, in  Ref. \cite{F} a complexified version of (\ref{ssd}) was considered.}
Our objective in this section is to demonstrate that (\ref{ssd}) comes about naturally if
one applies the method of nonlinear realizations to $OSp(1|2)$ supergroup and keeps the number of independent Goldstone superfields to a minimum.

Consider a real superspace $\mathcal{R}^{1|1}$ parametrized by a bosonic coordinate $t$ and a fermionic coordinate $\theta$, $\theta^2=0$. The supersymmetry transformations
\be\label{n1susy}
t'=t+a; \qquad \quad t'=t+i \epsilon\theta, \qquad \theta'=\theta+\epsilon,
\ee
where $a$ and $\epsilon$ are even and odd real supernumbers, respectively, realise the action of the $d=1$, $\mathcal{N}=1$ supersymmetry algebra
\be\label{n1algebra}
\{q,q\}=2 h
\ee
in the superspace.

Within the method of nonlinear realizations, $\mathcal{R}^{1|1}$ is represented by the supergroup element
\be
{\tilde g}=e^{i t h} e^{\theta q},
\ee
while the left action of the supergroup on itself, ${\tilde g}'=e^{i a h}e^{\epsilon q} \cdot {\tilde g}$, reproduces (\ref{n1susy}).  The covariant derivative, which anticommutes with the supersymmetry generator, reads
\be\label{cd}
\mathcal{D}=\partial_\theta-i\theta \partial_t, \qquad \mathcal{D}^2=-i \partial_t,
\ee
where $\partial_t=\frac{\partial}{\partial t}$, $\partial_\theta=\frac{\vec{\partial}}{\partial \theta}$.

Real bosonic and fermionic superfields are power series in $\theta$
\be
\rho(t,\theta)=b(t)+i\theta f(t), \qquad \psi(t,\theta)=F(t)+\theta B(t), \qquad
{\left(\mathcal{D} \rho\right)}^{*}=-\mathcal{D} \rho, \qquad {\left(\mathcal{D} \psi\right)}^{*}=\mathcal{D} \psi,
\ee
which involve the bosonic components $(b(t),B(t))$ and their fermionic partners $(f(t),F(t))$. The covariant derivative (\ref{cd}) and a real fermionic superfield $\psi(t,\theta)$ are the building blocks entering Eq. (\ref{ssd}) above.

In order to derive the $\mathcal{N}=1$ super--Schwarzian derivative within the method of nonlinear realizations, let us consider $osp(1|2)$ superconformal algebra
\begin{align}\label{salgebra}
&
[P,D]=i P, && [P,K]=2iD,
\nonumber\\[2pt]
&
[D,K]=i K, &&
[D,Q]=-\frac{i}{2} Q,
\nonumber\\[2pt]
&
[D,S]=\frac{i}{2} S, && [P,S]=-i Q,
\nonumber\\[2pt]
&
[K,Q]=i S, &&  \{Q,S\}=-2D,
\nonumber\\[4pt]
&
\{Q,Q\}=2 P, &&
\{S,S\}=2 K,
\end{align}
where $(P,D,K)$ are the bosonic generators of translations, dilatations and special conformal transformations, respectively. $Q$ and $S$ are the fermionic generators of supersymmetry transformations and superconformal boosts.

As the next step, each generator in the superalgebra (\ref{salgebra}) is accompanied by a real Goldstone superfield of the same Grassmann parity
and both $\mathcal{R}^{1|1}$ and the superfields on it are represented by the element
\be\label{g}
\tilde g=e^{i t h} e^{\theta q} e^{i\rho(t,\theta) P} e^{\psi(t,\theta) Q} e^{\phi(t,\theta) S} e^{i \mu(t,\theta) K} e^{i \nu(t,\theta) D}.
\ee
It is assumed that $(h,q)$ (anti)commute with $(P,D,K,Q,S)$.

Left multiplication by a group element $\tilde g'=g \cdot \tilde g$, where
$g=e^{i a P} e^{\epsilon Q} e^{\sigma S} e^{i c K} e^{i b D}$ involves real bosonic parameters $(a,b,c)$  and real fermionic parameters $(\epsilon,\sigma)$,
determines the action of the superconformal group $OSp(1|2)$ on the superfields. Focusing on the infinitesimal transformations and
making use of the Baker--Campbell--Hausdorff formula
\be\label{ser}
e^{iA}~ T~ e^{-iA}=T+\sum_{n=1}^\infty\frac{i^n}{n!}
\underbrace{[A,[A, \dots [A,T] \dots]]}_{n~\rm times},
\ee
one gets
\begin{align}\label{tr1}
&
\rho'=\rho+a; && &&
\nonumber\\[10pt]
&
\rho'=\rho+b\rho, && \nu'=\nu+b, &&
\nonumber\\[2pt]
&
\mu'=\mu-b\mu, &&
\psi'=\psi+\frac 12 b \psi, && \phi'=\phi-\frac 12 b\phi; &&
\nonumber\\[10pt]
&
\rho'=\rho+c\rho^2, && \nu'=\nu+2c\rho, &&
\nonumber\\[2pt]
&
\mu'=\mu+c-2c\rho\mu-ic\psi\phi, &&
\psi'=\psi+c\rho\psi, && \phi'=\phi-c\psi-c\rho\phi; &&
\nonumber\\[10pt]
&
\rho'=\rho+i\epsilon \psi, && \psi'=\psi+\epsilon; &&
\nonumber\\[10pt]
&
\rho'=\rho-i\sigma \rho \psi, && \nu'=\nu-2 i \sigma \psi, &&
\nonumber\\[2pt]
&
\mu'=\mu+i \sigma\phi+2i\sigma\mu\psi, &&
\psi'=\psi-\sigma \rho, && \phi'=\phi+\sigma+i\sigma\psi\phi. &&
\end{align}
Note that both the original and transformed superfields depend on the same arguments $(t,\theta)$ such that the transformations affect the form of the superfields only, e.g. $\delta \rho=\rho'(t,\theta)-\rho(t,\theta)$.
Computing the algebra of the infinitesimal transformations (\ref{tr1}), one can verify that it does reproduce the structure relations (\ref{salgebra}).\footnote{In order to verify the structure relations (\ref{salgebra}), one first computes the commutators $[\delta_1,\delta_2]=\delta_3$ acting upon $(\rho,\mu,\nu,\psi,\phi)$
for all the transformations entering (\ref{tr1}). Then one represents each $\delta$ as the product of a parameter and the corresponding generator, e.g. $\delta_a=a \cdot P$. Finally, one substitutes these into  $[\delta_1,\delta_2]=\delta_3$ and discards the parameters on both the left and right hand sides of the equality. For the case at hand, this yields (\ref{salgebra}) after the rescaling $(P,K,D)\to(iP,iK,iD)$.}

As the next step, one computes the invariant superfield combinations
\bea\label{inv}
&&
\omega_P=\left(\mathcal{D}\rho+i\psi \mathcal{D}\psi \right)e^{-\nu},
\nonumber\\[6pt]
&&
\omega_D=\mathcal{D}\nu-2i\phi\mathcal{D}\psi-2\mu \left(\mathcal{D}\rho+i\psi \mathcal{D}\psi \right),
\nonumber\\[6pt]
&&
\omega_K=\left(\mathcal{D}\mu+i\phi\mathcal{D}\phi+2i\mu\phi\mathcal{D}\psi+\mu^2(\mathcal{D}\rho+i\psi \mathcal{D}\psi)\right)e^{\nu},
\nonumber\\[6pt]
&&
\omega_Q=\left(\mathcal{D}\psi-\phi (\mathcal{D}\rho+i\psi \mathcal{D}\psi) \right)e^{-\frac{\nu}{2}},
\nonumber\\[6pt]
&&
\omega_S=\left(\mathcal{D}\phi+\mu (\mathcal{D}\psi-\phi(\mathcal{D}\rho+i\psi \mathcal{D}\psi)) \right)e^{\frac{\nu}{2}},
\eea
which originate from
\be
\tilde g^{-1} \mathcal{D} \tilde g=i \omega_P P+i \omega_D D +i \omega_K K+\omega_Q Q+\omega_S S+q.
\ee
Note that the Grassmann parities of the invariants (\ref{inv}) are opposite to those associated with the conventional Maurer--Cartan one--forms $\tilde g^{-1} d \tilde g$ because $\mathcal{D}$ is an odd operator.
When obtaining (\ref{inv}), the identities exposed in Appendix were heavily used.

At this stage, one can use the invariants (\ref{inv}) so as to impose constraints enabling one to eliminate some of the Goldstone superfields. By analogy with our study of the Schwarzian derivative in \cite{G1},
let us decide to keep the number of independent Goldstone superfields to a minimum and impose four
conditions
\be\label{const}
\omega_P=0, \qquad \omega_D=0, \qquad \omega_Q=g^{-1}, \qquad \omega_S=p,
\ee
where $g$ and $p$ are even supernumbers. It seems quite natural to set the fermionic invariants to zero and to demand the bosonic invariants to take constant values as only the $c$--numbers are observable.

The leftmost equation in (\ref{const}) gives
\be\label{c1}
\mathcal{D}\rho+i\psi \mathcal{D}\psi=0,
\ee
while the rest links $(\nu,\mu,\phi)$ to $\psi$
\be\label{sol}
e^{\frac{\nu}{2}}=g \mathcal{D}\psi, \qquad \phi=-\frac{\partial_t \psi}{{(\mathcal{D}\psi )}^2 }, \qquad \mu=\frac{p}{g {(\mathcal{D}\psi)}^2 }+\frac{1}{\mathcal{D}\psi} \mathcal{D}\left(\frac{\partial_t \psi}{{(\mathcal{D}\psi)}^2} \right).
\ee
Substituting these relations into the last remaining invariant $\omega_K$, one gets
\be
\omega_K=i g^2 \left(\frac{\mathcal{D}^4 \psi}{{\mathcal{D}\psi}}-2 \frac{\mathcal{D}^3 \psi}{{\mathcal{D}\psi}} \frac{\mathcal{D}^2 \psi}{{\mathcal{D}\psi}} \right).
\ee
Up to an irrelevant constant factor, this coincides with the $\mathcal{N}=1$ super--Schwarzian derivative in (\ref{ssd}).

We conclude this section with a discussion of symmetries of (\ref{ssd}). If one is interested in infinitesimal transformations, it suffices to consider (\ref{tr1}) and focus on the transformation laws of $\rho$ and $\psi$. A straightforward computation shows that both (\ref{ssd}) and the supplementary condition (\ref{c1}) hold invariant.

If one is concerned with finite transformations, then, following Ref. \cite{F}, one has to consider a generic super--diffeomorphism of $\mathcal{R}^{1|1}$
\be\label{c11}
t'=\rho(t,\theta), \qquad \theta'=\psi(t,\theta),
\ee
under which the covariant derivative transforms homogeneously. The condition
\be\label{c12}
\mathcal{D}=\left(\mathcal{D}\theta'\right) \mathcal{D}'.
\ee
yields the restriction on the bosonic superfield $\rho$
\be\label{CON}
\mathcal{D}\rho+i\psi \mathcal{D}\psi=0,
\ee
yet leaves the fermionic superfield $\psi$ unconstrained. Acting on (\ref{CON}) by the covariant derivative $\mathcal{D}$, one gets the equation
\be\label{COND1}
\partial_t \rho-i\psi \partial_t \psi-{( \mathcal{D} \psi )}^2=0,
\ee
which can be used to fix $\rho$ provided $\psi$ is known. Note that within the method of nonlinear realizations the supplementary condition (\ref{CON}) comes about as the constraint $\omega_P=0$.

Given $\rho(t,\theta)$ and $\psi(t,\theta)$ obeying (\ref{CON}), consider a coordinate transformation (\ref{c11}) and a new real fermionic superfield $\psi'(t',\theta')=\psi'(\rho,\psi)$. Taking into account Eq. (\ref{COND1}), one gets the formula
\be
\partial_t=(\partial_t \theta') \mathcal{D}'+{\left(\mathcal{D} \theta'\right)}^2 \partial_{t'},
\ee
which enables one to obtain the transformation law of the $\mathcal{N}=1$ super--Schwarzian\footnote{It proves helpful to keep in mind the identities
$\frac{\mathcal{D}^4 \psi}{{\mathcal{D}\psi}}-2 \frac{\mathcal{D}^3 \psi}{{\mathcal{D}\psi}} \frac{\mathcal{D}^2 \psi}{{\mathcal{D}\psi}}=\mathcal{D}\psi \mathcal{D}\left(\frac{\mathcal{D}^3 \psi}{{\mathcal{D}\psi } {\mathcal{D}\psi} }\right)=\mathcal{D}\psi \mathcal{D}^2 \left(\frac{{\mathcal{D}^2 \psi}}{{\mathcal{D} \psi} {\mathcal{D}\psi} }\right)
=-{\mathcal{D} \psi} \mathcal{D}^3 \left( \frac{1}{{\mathcal{D}\psi}}\right)$.}
\be
S[\psi'(\rho,\psi);t,\theta]=S[\psi(t,\theta);t,\theta]+{\left(\mathcal{D} \theta' \right)}^3 S[\psi'(t',\theta');t',\theta'].
\ee
Thus, if $S[\psi'(t',\theta');t',\theta']=0$, the $\mathcal{N}=1$ super--Schwarzian derivative holds invariant under the change of the argument $\psi(t,\theta) \to \psi'(\rho(t,\theta),\psi(t,\theta))$. Solving $S[\psi'(t',\theta');t',\theta']=0$ and integrating the analogue of Eq. (\ref{COND1}), one finally gets
\be\label{TRF}
\psi'=\alpha+\frac{\beta+\psi}{c \rho+d}, \qquad \rho'=\frac{a\rho+b}{c\rho+d}-\frac{i(\psi\psi'-\alpha\beta)}{c\rho+d},
\ee
where $(a,b,c,d)$ are real even supernumbers obeying $ad-cb=1$ and $(\alpha,\beta)$ are real odd supernumbers.

Eq. (\ref{TRF}) describes the finite form of $OSp(1|2)$ transformations acting upon the form of the superfields $\rho$ and $\psi$ which leave the $\mathcal{N}=1$ super--Schwarzian and the supplementary condition (\ref{CON}) invariant (cf. (\ref{tr1})).

\vspace{0.5cm}

\noindent
{\bf 3. $\mathcal{N}=2$ super--Schwarzian derivative via nonlinear realizations}\\

\noindent
The $\mathcal{N}=2$ super--Schwarzian derivative \cite{Cohn}
\be\label{n2ssd}
S[\psi(t,\theta,\bar\theta);t,\theta,\bar\theta]=\frac{\partial_t \mathcal{D}\psi}{\mathcal{D}\psi}-\frac{\partial_t \bar{\mathcal{D}}\bar\psi}{\bar{\mathcal{D}}\bar\psi}-2i \frac{\partial_t \psi \partial_t \bar\psi}{\mathcal{D}\psi \bar{\mathcal{D}}\bar\psi}
\ee
involves a complex chiral fermionic superfield $\psi$ defined on $\mathcal{R}^{1|2}$ superspace and its complex conjugate partner $\bar\psi=\psi^{*}$
\be
\bar{\mathcal{D}}\psi=0, \qquad \mathcal{D}\bar\psi=0.
\ee
$\mathcal{R}^{1|2}$
is parametrized by a bosonic coordinate $t$ and a pair of complex conjugate fermionic coordinates $(\theta,\bar\theta)$, $\theta\bar\theta=-\bar\theta \theta$, $\theta^2=\bar\theta^2=0$.
The $d=1$, $\mathcal{N}=2$ supersymmetry transformations read
\be\label{n2susy}
t'=t+a;  \qquad \theta'=\theta+\epsilon, \qquad \bar\theta'=\bar\theta+\bar\epsilon, \qquad t'=t+i (\epsilon\bar\theta+\bar\epsilon \theta),
\ee
and the covariant derivatives, which anticommute with the supersymmetry generators, are realised as follows:
\be\label{cd2}
\mathcal{D}=\partial_\theta-i\bar\theta \partial_t, \qquad \bar{\mathcal{D}}=\partial_{\bar\theta}-i \theta \partial_t,
\qquad
\{\mathcal{D},\bar{\mathcal{D}} \}=-2i\partial_t, \qquad \mathcal{D}^2={\bar{\mathcal{D}}}^2=0,
\ee
where $\partial_t=\frac{\partial}{\partial t}$, $\partial_\theta=\frac{\vec{\partial}}{\partial \theta}$, $\partial_{\bar\theta}=\frac{\vec{\partial}}{\partial \bar\theta}$.

Within the method of nonlinear realizations $\mathcal{R}^{1|2}$ is identified with the supergroup element
\be
{\tilde g}=e^{i t h} e^{\theta q+\bar\theta \bar q}, \qquad \{q,{\bar q} \}=2 h,
\ee
while (\ref{n2susy}) follow from ${\tilde g}'=e^{i a h}e^{\epsilon q+\bar\epsilon \bar q} \cdot {\tilde g}$.

$\mathcal{N}=2$ superfields are power series in $\theta$ and $\bar\theta$ which involve component fields dependent on $t$. If $\rho$ is a real bosonic superfield and $\psi$ is a complex fermionic superfield, the complex
conjugation rules for their covariant derivatives read
\be
 {\left(\mathcal{D}\rho\right)}^{*}=-\bar{\mathcal{D}} \rho, \qquad {\left(\mathcal{D}\psi\right)}^{*}=\bar{\mathcal{D}}\bar\psi, \qquad {\left(\bar{\mathcal{D}}\psi\right)}^{*}=\mathcal{D}\bar\psi.
\ee

In order to obtain the $\mathcal{N}=2$ super--Schwarzian derivative within the method of nonlinear realizations, let us consider $su(1,1|1)$ superconformal algebra
\begin{align}\label{salgebra2}
&
[P,D]=i P, && [P,K]=2iD,
\nonumber\\[2pt]
&
[D,K]=i K, &&
[D,Q]=-\frac{i}{2} Q,
\nonumber\\[2pt]
&
[D,\bar Q]=-\frac{i}{2} \bar Q, && [D,S]=\frac{i}{2} S
\nonumber\\[2pt]
&
[D,\bar S]=\frac{i}{2} \bar S, && [P,S]=-i Q,
\nonumber\\[2pt]
&
[P,\bar S]=-i \bar Q, && [K,Q]=i S,
\nonumber\\[2pt]
&
[K,\bar Q]=i \bar S, && [J,Q]=\frac 12 Q,
\nonumber\\[2pt]
&
[J,\bar Q]=-\frac 12 \bar Q, && [J,S]=\frac 12 S,
\nonumber\\[2pt]
&
[J,\bar S]=-\frac 12 \bar S, && \{Q,\bar Q\}=2 P,
\nonumber\\[2pt]
&
\{Q,\bar S\}=-2(D+i J), && \{S,\bar S\}=2 K,
\nonumber\\[6pt]
&
\{\bar Q,S\}=-2(D-i J), &&
\end{align}
where $(P,D,K,J)$ are bosonic generators and $(Q,\bar Q,S,\bar S)$ are their fermionic partners. As compared to the $\mathcal{N}=1$ case, there appears a new $U(1)$--symmetry generator $J$, while $Q$ and $S$ become complex.

Before constructing a group--theoretic element similar to (\ref{g}), it proves instructive to make recourse to a generic super--diffeomorphism of $\mathcal{R}^{1|2}$
\be\label{sdn2}
t'=\rho(t,\theta,\bar\theta), \qquad \theta'=\psi(t,\theta,\bar\theta), \qquad \bar\theta'=\bar\psi(t,\theta,\bar\theta),
\ee
where $\rho$ is a real bosonic superfield and $\psi$ is a complex fermionic superfield,
and find conditions which follow from the requirement that the covariant derivatives transform homogeneously. The elementary computation
\be
\mathcal{D}=\left(\mathcal{D} t'+i\bar\theta' \mathcal{D}\theta'\right) \partial_{t'}+\left(\mathcal{D}\theta'\right) \mathcal{D}'+\left(\mathcal{D}\bar\theta'\right) \partial_{\bar\theta'}=(\mathcal{D}\theta') \mathcal{D}' \ee
gives
\be\label{c22}
\mathcal{D}\bar\psi=0, \quad \mathcal{D}\rho+i\bar\psi \mathcal{D}\psi=0,
\ee
while $\bar{\mathcal{D}}=\left(\bar{\mathcal{D}}{\bar\theta}'\right) \bar{\mathcal{D}}'$ yields the complex conjugate restrictions
\be\label{c23}
\bar{\mathcal{D}}\psi=0, \quad \bar{\mathcal{D}}\rho+i\psi \bar{\mathcal{D}}\bar\psi=0.
\ee
These are the $\mathcal{N}=2$ analogues of Eq. (\ref{CON}) above. Note the simple corollary of Eqs. (\ref{c22}) and (\ref{c23})
\be\label{c24}
\partial_t \rho=(\mathcal{D}\psi) (\bar{\mathcal{D}}\bar\psi)+i\psi \partial_t \bar\psi-i \partial_t \psi \bar\psi,
\ee
which may be used to fix $\rho$ provided $\psi$ is known.

In what follows, we shall assume that the conditions (\ref{c22}) and  (\ref{c23}) hold. As a matter of fact, the method of nonlinear realizations allows one to reproduce the equation for $\rho$, which comes about as the constraint $\omega_P$=0, but not the chirality condition for the fermionic superfield $\psi$.

In view of all the foregoing, consider the group--theoretic element
\bea\label{gn2}
&&
\tilde g=e^{i t h} e^{\theta q+\bar\theta \bar q} e^{i\rho P} e^{\psi Q+\bar\psi \bar Q} e^{\phi S+\bar\phi \bar S} e^{i \mu  K} e^{i \nu D} e^{i \lambda J},
\eea
where $(\rho,\mu,\nu,\lambda)$ are real bosonic superfields and $(\psi,\phi)$ are complex fermionic superfields.
Note that such a choice of $\tilde g$ is suggested by the previous study of $d=1$, $\mathcal{N}=2$ superconformal mechanics within the method of nonlinear realizations \cite{IKL}.
The superconformal invariants, which derive from ${\tilde g}^{-1} \mathcal{D} \tilde g$, read
\bea\label{inv2}
&&
\omega_D=\mathcal{D}\nu-2i \bar\phi \mathcal{D}\psi,
\nonumber\\[2pt]
&&
\omega_K=\left(\mathcal{D} \mu+i\phi \mathcal{D} \bar\phi+i\bar\phi \mathcal{D} \phi+2i\mu \bar\phi \mathcal{D}\psi  \right)e^{\nu},
\nonumber\\[2pt]
&&
\omega_J=\mathcal{D}\lambda+2\bar\phi \mathcal{D}\psi,
\nonumber\\[2pt]
&&
\omega_Q=\mathcal{D}\psi e^{-\frac{\nu}{2}}e^{-\frac{i\lambda}{2}},
\nonumber\\[2pt]
&&
\omega_S=\left(\mathcal{D}\phi+\mu \mathcal{D}\psi-i\phi\bar\phi\mathcal{D}\psi \right) e^{\frac{\nu}{2}}e^{-\frac{i\lambda}{2}},
\nonumber\\[2pt]
&&
\omega_{\bar S}=\mathcal{D}\bar\phi e^{\frac{\nu}{2}}e^{\frac{i\lambda}{2}}.
\eea
When obtaining Eqs. (\ref{inv2}), the identities gathered in Appendix were heavily used.

As the next step, let us impose constraints similar to those in the preceding section
\be\label{constr2}
\omega_D=0, \qquad \omega_Q=g^{-1}, \qquad \omega_S=p,
\ee
where $g$ and $p$ are complex even supernumbers. They allow one to express $(\mu,\nu,\lambda,\phi)$ in terms of $\psi$
\bea\label{for}
&&
\mu=\frac{pg}{e^{\nu}}+i\phi\bar\phi-\frac{\mathcal{D}\phi}{\mathcal{D}\psi}, \qquad
e^{\nu}=g \bar g \mathcal{D}\psi \bar{\mathcal{D}} \bar\psi, \qquad
e^{i\lambda}=\frac{g \mathcal{D}\psi}{\bar g \bar{\mathcal{D}} \bar\psi },  \qquad \phi=-\frac{\partial_t \psi}{\mathcal{D}\psi \bar{\mathcal{D}} \bar\psi},
\eea
which, in their turn, ensure $\phi$ to be a chiral superfield, $\bar{\mathcal{D}} \phi=0$, and force the remaining invariants to vanish, $\omega_K=\omega_J=\omega_{\bar S}=0$.
Finally, taking into account that $\mu$ is a real superfield, $\bar\mu=\mu$, one gets the equation
\be\label{SSDn2}
\frac{\partial_t \mathcal{D}\psi}{\mathcal{D}\psi}-\frac{\partial_t \bar{\mathcal{D}}\bar\psi}{\bar{\mathcal{D}}\bar\psi}-2i \frac{\partial_t \psi \partial_t \bar\psi}{\mathcal{D}\psi \bar{\mathcal{D}}\bar\psi}=\frac{\bar p \bar g-p g}{g \bar g},
\ee
the right hand side of which reproduces the $\mathcal{N}=2$ super--Schwarzain derivative (\ref{n2ssd}).

As compared to the $\mathcal{N}=1$ case, the constraints (\ref{constr2}) turn out to be more stringent and result in a variant of $\mathcal{N}=2$ super--Schwarzain mechanics in which the super--Schwarzian derivative is equal to a (coupling) constant $\frac{\bar p \bar g-p g}{g \bar g}$. The latter is an $\mathcal{N}=2$ analogue of the model studied recently in \cite{G2}. As was mentioned in the Introduction, our primarily concern in this work is to understand how the super--Schwarzian derivatives may be obtained within the method of nonlinear realizations. The dynamics of the specific model (\ref{SSDn2}) will not be studied any further.

Concluding this section, let us discuss symmetries of Eqs. (\ref{n2ssd}) and (\ref{c22}), (\ref{c23}).
The infinitesimal form of such transformations follows from
\be
\tilde g'=e^{i a P} e^{\epsilon Q+\bar\epsilon \bar Q} e^{\sigma S+\bar\sigma \bar S} e^{i c K} e^{i b D} e^{i \xi J} \cdot \tilde g,
\ee
where $\tilde g$ is given in (\ref{gn2}), while $(a,b,c,\xi)$ and $(\epsilon,\sigma)$ are bosonic and fermionic infinitesimal parameters, respectively. Implementing the Baker--Campbell--Hausdorff formula (\ref{ser}) and discarding the transformation laws of $(\mu,\nu,\lambda,\phi,\bar\phi)$, one gets
\begin{align}\label{tr3}
&
\rho'=\rho+a, && \psi'=\psi, && \bar\psi'=\bar\psi;
\nonumber\\[6pt]
&
\rho'=\rho+b\rho, && \psi'=\psi+\frac 12 b \psi, && \bar\psi'=\bar\psi+\frac 12 b \bar\psi;
\nonumber\\[6pt]
&
\rho'=\rho+c \rho^2, && \psi'=\psi+c \rho \psi, && \bar\psi'=\bar\psi+c\rho \bar\psi;
\nonumber\\[6pt]
&
\rho'=\rho, && \psi'=\psi+\frac{i}{2} \xi \psi, && \bar\psi'=\bar\psi-\frac{i}{2} \xi \bar\psi;
\nonumber\\[6pt]
&
\rho'=\rho+i\left(\epsilon\bar\psi+\bar\epsilon \psi \right), && \psi'=\psi+\epsilon, && \bar\psi'=\bar\psi+\bar\epsilon;
\nonumber\\[6pt]
&
\rho'=\rho-i\rho \left(\sigma\bar\psi+\bar\sigma\psi \right), && \psi'=\psi-\rho \sigma+i\sigma \psi\bar\psi, && \bar\psi'=\bar\psi-\rho \bar\sigma-i\bar\sigma \psi\bar\psi.
\end{align}
As in the $\mathcal{N}=1$ case, the transformations act upon the form of the superfields only and do not affect the arguments $(t,\theta,\bar\theta)$. It is easy to compute the commutators $[\delta_1,\delta_2]$ and verify that they do reproduce the structure relations (\ref{salgebra2}).\footnote{As in the $\mathcal{N}=1$, after computing the algebra one has to rescale the bosonic generators $(P,D,K,J)\to(iP,iD,iK,iJ)$ so as to fit the notation in (\ref{salgebra2}).} A straightforward calculation then shows that both (\ref{n2ssd}) and the supplementary conditions (\ref{c22}), (\ref{c23}) hold invariant under the infinitesimal transformations.

In order to determine a finite form of symmetries of the $\mathcal{N}=2$ super--Schwarzian, one considers the coordinate transformation (\ref{sdn2}), which obeys the subsidiary conditions (\ref{c22}), (\ref{c23}), and a new complex  chiral fermionic superfield $\psi'(t',\theta',\bar\theta')=\psi'(\rho,\psi,\bar\psi)$. Taking into account the relations
\be
\mathcal{D}=\left(\mathcal{D}\theta'\right) \mathcal{D}', \qquad
\bar{\mathcal{D}}=\left(\bar{\mathcal{D}}{\bar\theta}'\right) \bar{\mathcal{D}}', \qquad \partial_t=(\partial_t \theta') \mathcal{D}'+(\partial_t \bar\theta') \bar{\mathcal{D}}'+(\mathcal{D}\theta' \bar{\mathcal{D}} \bar\theta') \partial_{t'},
\ee
one gets
\be
S[\psi'(\rho,\psi,\bar\psi);t,\theta,\bar\theta]=S[\psi(t,\theta,\bar\theta);t,\theta,\bar\theta]+(\mathcal{D}\theta' \bar{\mathcal{D}} \bar\theta') S[\psi'(t',\theta',\bar\theta');t',\theta',\bar\theta'].
\ee
Thus, the $\mathcal{N}=2$ super--Schwarzian derivative remains intact under the change of the argument $\psi(t,\theta,\bar\theta)\to \psi'(\rho(t,\theta,\bar\theta),\psi(t,\theta,\bar\theta),\bar\psi(t,\theta,\bar\theta))$, provided $S[\psi'(t',\theta',\bar\theta');t',\theta',\bar\theta']=0$. Solving the latter equation and integrating the analogue of (\ref{c24}) for $\rho'$, one gets
\bea\label{R1}
&&
\psi'=\alpha-\frac{\beta}{c\rho+d}+\psi \frac{e^{-i v}}{c\rho+d}\left(c-\frac{i\beta \bar\beta}{c\rho+d} \right)-\frac{i\psi\bar\psi c \beta}{{(c\rho+d)}^2},
\nonumber\\[2pt]
&&
\rho'=\frac{a\rho+b}{c\rho+d}-\frac{i(\alpha\bar\beta-\beta\bar\alpha)}{c\rho+d}-\frac{i\psi e^{-iv}}{c\rho+d} \left(c\bar\alpha-\frac{\bar\beta(c-i\beta\bar\alpha)}{c\rho+d} \right)
\nonumber\\[2pt]
&&
\qquad
-\frac{i\bar\psi e^{iv}}{c\rho+d} \left(c\alpha-\frac{\beta(c+i\alpha\bar\beta)}{c\rho+d} \right)-\frac{\psi\bar\psi c}{{(c\rho+d)}^2} \left(\alpha\bar\beta+\beta\bar\alpha-\frac{2\beta\bar\beta}{c\rho+d} \right),
\eea
where $(a,b,c,d,v)$ are real even supernumbers obeying $ad-cb=c^2$ and $(\alpha,\beta)$ are complex odd supernumbers.\footnote{The standard form of $SL(2,R)$ transformations with $ad-cb=1$ is recovered by rescaling $\psi'\to \frac{1}{c} \psi'$, $\rho'\to \frac{1}{c^2} \rho'$, $\frac{a}{c^2}\to a$, $\frac{b}{c^2}\to b$.} Eq. (\ref{R1}) describes the finite form of $SU(1,1|1)$ transformations exposed in (\ref{tr3}).
Finally, it is straightforward to verify that the supplementary conditions (\ref{c22}), (\ref{c23}) hold invariant under the transformation (\ref{R1}).

\vspace{0.5cm}
\noindent
{\bf 4. Discussion}\\

\noindent
To summarize, in this work we have demonstrated that the $\mathcal{N}=1$ and $\mathcal{N}=2$ super--Schwarzian derivatives can be obtained within the method of nonlinear realizations applied to $OSp(1|2)$ and $SU(1,1|1)$ superconformal groups, thus providing an alternative to the existing approaches.

Let us discuss possible further developments. Although it is not quite clear whether the construction in this work might tell us something new about superconformal field theory, it can definitely be used to generate super--Schwarzians invariant under a given supergroup. Such objects are indispensable for constructing $\mathcal{N}>2$ supersymmetric extensions of the Sachdev--Ye--Kitaev model. In this regard, the most pressing issue is to generalise the analysis above to the $\mathcal{N}=4$ case, i.e. to treat $SU(1,1|2)$ superconformal group in a similar fashion.  

In the literature there is some controversy on the latter point. In Ref. \cite{Sch} it is stated that an $\mathcal{N}=4$ super--Schwarzian is a non--local expression and only the covariant derivative of it is given in explicit form.
However, because the $R$-symmetry subalgebra in \cite{Sch} is $so(4)$, the case seems to correspond to $Osp(4|2)$ rather than $SU(1,1|2)$ (see also a related work \cite{GR}).
Mathematicians report an
obstruction to obtain a projective cocycle for $\mathcal{N} \geq 3$ (see, e.g., the discussion in \cite{MD}). An $\mathcal{N}=4$ super--Schwarzian proposed in \cite{AH} does not seem to be invariant under finite $SU(1,1|2)$ transformations (any super--Schwarzian should be a homogeneous function of degree zero under the rescaling $\psi\to b\psi$).

A preliminary consideration shows that an $\mathcal{N}=4$ super--Schwarzian
generated by the method of nonlinear realizations might read
\be\label{n4ss}
\frac{\mathcal{D}^\alpha \psi_\beta \partial_t \bar{\mathcal{D}}_\alpha  \bar\psi^\beta}{ {\mathcal{D} \psi \bar{\mathcal{D}} \bar\psi}} -\frac{\bar{\mathcal{D}}_\alpha \bar\psi^\beta \partial_t  \mathcal{D}^\alpha \psi_\beta}{{\mathcal{D} \psi \bar{\mathcal{D}} \bar\psi}},
\ee
where $\psi_\alpha$ is a fermionic chiral superfield on $\mathcal{R}^{1|4}$ superspace carrying an $SU(2)$ spinor index $\alpha=1,2$, $\bar\psi^\alpha$ is its complex conjugate ${(\psi_\alpha)}^{*}=\bar\psi^\alpha$, and
$\mathcal{D}^\alpha$, $\bar{\mathcal{D}}_\alpha$ are the covariant derivatives. Above we abbreviated $\left(\mathcal{D} \psi \bar{\mathcal{D}} \bar\psi\right)=\mathcal{D}^\alpha \psi_\beta \bar{\mathcal{D}}_\alpha \bar\psi^\beta$. Yet, it turns out that along with (\ref{n4ss}) extra quadratic constraints on $\psi_\alpha$ appear, which still need to be understood. We hope to report on the progress as well as to describe a more general case of the $D(2,1;\alpha)$ super--Schwarzian elsewhere.

An elegant derivation of the $\mathcal{N}=1$ and $\mathcal{N}=2$ super--Schwarzian derivatives within the context of a one-dimensional $Osp(N|2M)$ pseudoparticle mechanics was proposed in \cite{AG}. It would be interesting to see if the analysis in \cite{AG} can be generalised to the $\mathcal{N}=4$ case, which should link to the $Osp(4|2)$ super--Schwarzian in \cite{Sch}.

A connection between the conventional second order conformal mechanics and the Schwar\-zian mechanics was discussed in a very recent work \cite{FM}. It would be interesting to explore whether the analysis in \cite{FM} can be extended to produce the super--Schwarzians. The construction of higher derivative superconformal mechanics of the Schwarzian type along the lines in \cite{MM} is of interest as well.
\vspace{0.5cm}

\noindent{\bf Acknowledgements}\\

\noindent
This work was supported by the Russian Foundation for Basic Research, grant No 20-52-12003.

\vspace{0.5cm}

\noindent
{\bf Appendix}

\vspace{0.5cm}

\noindent
In this Appendix we gather some identities which were used in the main text when computing the superconformal invariants (\ref{inv}) and (\ref{inv2}).

The identities which facilitate the derivation of Eq. (\ref{inv}) read
\begin{align}
&
e^{-i\nu D} P e^{i\nu D}=e^{-\nu} P, && e^{-i\nu D} K e^{i\nu D}=e^{\nu} K,
\nonumber\\[2pt]
&
e^{-i\nu D} Q e^{i\nu D}=e^{-\frac{\nu}{2}} Q, &&
e^{-i\nu D} S e^{i\nu D}=e^{\frac{\nu}{2}} S,
\nonumber\\[2pt]
&
e^{-i\mu K} P e^{i\mu K}=P-2\mu D+\mu^2 K, && e^{-i\mu K} D e^{i\mu K}=D-\mu K,
\nonumber
\end{align}
\begin{align}
&
e^{-i\mu K} Q e^{i\mu K}=Q+\mu S, && e^{-i\rho P} D e^{i\rho P}=D+\rho P,
\nonumber\\[2pt]
&
e^{-i\rho P} K e^{i\rho P}=K+2\rho D+\rho^2 P, && e^{-i\rho P} S e^{i\rho P}=S-\rho Q.
\nonumber
\end{align}
Note that these relations are also valid for the $su(1,1|1)$ superconformal algebra, in which case $Q$ and $S$ are regarded complex.
In that case the identities involving $\bar Q$, $\bar S$ follow by the Hermitian conjugation.

When computing the superconformal invariants (\ref{inv2}), the following identities:
\bea
&&
e^{-(\phi S+\bar\phi \bar S)} Q e^{\phi S+\bar\phi \bar S}=Q+2\bar\phi (D+iJ)-i\phi\bar\phi S,
\nonumber\\[2pt]
&&
e^{-(\psi Q+\bar\psi \bar Q)} \left( \mathcal{D} e^{\psi Q+\bar\psi \bar Q}\right)=\mathcal{D}\psi \left(Q-\bar\psi P \right),
\nonumber\\[2pt]
&&
e^{-(\phi S+\bar\phi \bar S)} P e^{\phi S+\bar\phi \bar S}=P-i\phi Q-i\bar\phi\bar Q+2\phi\bar\phi J,
\nonumber\\[2pt]
&&
e^{-(\psi Q+\bar\psi \bar Q)} \left( \bar{\mathcal{D}} e^{\psi Q+\bar\psi \bar Q}\right)=\bar{\mathcal{D}} \bar\psi \left(\bar Q-\psi P \right),
\nonumber\\[2pt]
&&
e^{-(\phi S+\bar\phi \bar S)} \bar Q e^{\phi S+\bar\phi \bar S}=\bar Q+2\phi (D-iJ)+i\phi\bar\phi \bar S,
\nonumber\\[2pt]
&&
e^{-(\phi S+\bar\phi \bar S)} \left( \mathcal{D} e^{\phi S+\bar\phi \bar S}\right)=\mathcal{D}\phi \left(S-\bar\phi K \right)+\mathcal{D} \bar\phi \left(\bar S-\phi K \right),
\nonumber\\[2pt]
&&
e^{-(\phi S+\bar\phi \bar S)} \left( \bar{\mathcal{D}} e^{\phi S+\bar\phi \bar S}\right)=\bar{\mathcal{D}} \bar\phi \left(\bar S-\phi K \right)+\bar{\mathcal{D}} \phi \left(S-\bar\phi K\right),
\nonumber\\[2pt]
&&
e^{-i\lambda J} Q e^{i\lambda J}=e^{-\frac{i\lambda}{2}} Q, \qquad
e^{-i\lambda J} \bar Q e^{i\lambda J}=e^{\frac{i\lambda}{2}} \bar Q,
\nonumber\\[2pt]
&&
e^{-i\lambda J} S e^{i\lambda J}=e^{-\frac{i\lambda}{2}} S, \qquad
e^{-i\lambda J} \bar S e^{i\lambda J}=e^{\frac{i\lambda}{2}} \bar S,
\nonumber
\eea
proved helpful. Recall that $\psi$ is a chiral fermionic superfield.


\vspace{0.5cm}

\begin{thebibliography}{nn}
\bibitem{FGMS}
W. Fu, D. Gaiotto, J. Maldacena, S. Sachdev, {\it Supersymmetric
Sachdev--Ye--Kitaev models}, Phys. Rev. D {\bf 95} (2017) 026009,  arXiv:1610.08917.
\bibitem{BBN}
M. Berkooz, N. Brukner, V. Narovlansky, A. Raz, {\it The double scaled limit of super--symmetric SYK models}, arXiv:2003.04405.
\bibitem{F}
D. Friedan, {\it Notes on string theory and two--dimensional conformal field theory}, Unified String Theories: proceedings (ed. by M.B. Green and D.J. Gross). Singapore, World Scientific, 1985.
\bibitem{Cohn}
J.D. Cohn, {\it $\mathcal{N}=2$ super--Riemann surfaces}, Nucl. Phys. B {\bf 284} (1987) 349.
\bibitem{Sch}
K. Schoutens, {\it $O(n)$ extended superconformal field theory in superspace}, Nucl. Phys. B {\bf 295} (1988) 634.
\bibitem{MD}
J.P. Michel, C. Duval, {\it On the projective geometry of the supercircle: a unified construction of the super cross--ratio and Schwarzian derivative}, Int. Math. Res. Not. 2008 (2008) 054, arXiv:0710.1544.
\bibitem{CWZ}
S.R. Coleman, J. Wess, B. Zumino, {\it Structure of phenomenological Lagrangians. I}, Phys. Rev. {\bf 177} (1969) 2239.
\bibitem{IO}
E.A. Ivanov, V.I. Ogievetsky, {\it The inverse Higgs phenomenon in
nonlinear realizations}, Theor. Math. Phys. {\bf 25} (1975) 1050.
\bibitem{G1}
A. Galajinsky, {\it Schwarzian mechanics via nonlinear realizations}, Phys. Lett. B {\bf 795} (2019) 277, arXiv:1905.01935.
\bibitem{IKL}
E. Ivanov, S. Krivonos, V. Leviant, {\it Geometric superfield approach to superconformal mechanics}, J. Phys. A {\bf 22} (1989) 4201.
\bibitem{G2}
A. Galajinsky, {\it A variant of Schwarzian mechanics}, Nucl. Phys. B {\bf 936} (2018) 661, arXiv:1809.00904.
\bibitem{GR}
S.J. Gates, Jr., L. Rana, {\it A proposal for $\mathcal{N}_0$ extended supersymmetry in integrable systems}, Phys. Lett. B {\bf 369} (1996) 269, hep-th/9510152.
\bibitem{AH}
S. Aoyama, Y. Honda, {\it $\mathcal{N}=4$ super--Schwarzian theory on the coadjoint orbit
and $PSU(1,1|2)$}, JHEP {\bf 06} (2018) 070, arXiv:1801.06800.
\bibitem{AG}
K.M. Apfeldorf, J. Gomis, {\it Superconformal theories from pseudoparticle mechanics}, Nucl. Phys. B {\bf 411} (1994) 745, hep-th/9303085.
\bibitem{FM}
S. Filyukov, I. Masterov, {\it On the Schwarzian counterparts of conformal mechanics}, arXiv:2004.03304.
\bibitem{MM}
I. Masterov, B. Merzlikin, {\it Superfield approach to higher derivative $\mathcal{N}=1$ superconformal mechanics}, JHEP {\bf 1911} (2019) 165, arXiv:1909.12574.
\end{thebibliography}
\end{document}